\documentclass[superscriptaddress,prb,aps,10pt,twocolumn]{revtex4}
\usepackage{graphicx}
\usepackage{graphics}
\usepackage{epsfig}
\usepackage{amsmath}
\usepackage{amssymb}
\usepackage{epstopdf}
\usepackage{subfigure}
\begin{document}

\def \etal{{\it et al}.~}
\def \etaln{{\it et al}.}

\title{Preparation of Metal Mixed Plastic Superconductors: Electrical  Properties of Tin-Antimony Thin Films on Plastic Substrates}

\author{Andrew P. Stephenson}
 \email{aps@physics.uq.edu.au}
 \affiliation{Centre for Organic Photonics and Electronics, School of Physical Sciences, University of Queensland, Brisbane QLD 4072, Australia}
\author{Ujjual Divakar} 
\affiliation{Centre for Organic Photonics and Electronics, School of Physical Sciences, University of Queensland, Brisbane QLD 4072, Australia}
\author{Adam P. Micolich}
 \affiliation{School of Physics, University of New South Wales, Sydney NSW 2052, Australia}
\author{Paul Meredith}
\affiliation{Centre for Organic Photonics and Electronics, School of Physical Sciences, University of Queensland, Brisbane QLD 4072, Australia}
\author{Ben J. Powell}
\affiliation{Centre for Organic Photonics and Electronics, School of Physical Sciences, University of Queensland, Brisbane QLD 4072, Australia}

\date{\today}

\begin{abstract}
Metal mixed polymers are a cheap and effective way to produce flexible metals and superconductors.  As part of an on-going effort to learn how to tune the properties of these systems with ion implantation, we present a study of the electrical properties of these systems prior to metal-mixing. We show that the electrical properties of tin-antimony thin films are remarkably robust to variations in the substrate morphology. We demonstrate that the optical absorbance of the films at a fixed wavelength provides a reliable and reproducible characterization of the relative film thickness. We find that as the film thickness is reduced, the superconducting transition in the unimplanted thin films is broadened, but the onset of the transition remains at $\sim$3.7~K, the transition temperature of bulk Sn. This is in marked contrast to the behavior of metal mixed films, which suggests that the metal mixing process has a significant effect of the physics of the superconducting state beyond that achieved by reducing the film thickness alone.

\end{abstract}

\maketitle

\clearpage

\section{Introduction}
The last twenty years has witnessed an explosion of interest in the electronic properties of organic materials.\cite{HeegerRMP,MacDiarmidRMP,ShirakawaRMP}  This interest is driven by their potential use in `soft electronics', which exploits properties of organic materials, such as low cost and mechanical flexibility, that are not typically found in traditional inorganic electronic materials.  Indeed, flexible organic displays and electronic devices are now beginning to penetrate the market, and future soft electronic materials will undoubtedly benefit from lower scaled costs and greater manufacturing simplicity.\cite{Voss2000}

While the main focus of research to date has been obtaining semiconducting and metallic organic materials, there is also a long history of research into superconducting organic materials.\cite{Powell2006}  Typically, organic superconductors are salts that form highly ordered crystals.  In these salts, electronic charge is transferred between an organic molecule [e.g., bis(ethylene-dithio)tetrathiafulvalene (BEDT-TTF), tetramethyl-tetraselenafulvalene (TMTSF) or buckminsterfullerene (C$_{60}$)] and a counter-ion, which is usually inorganic.\cite{Ishiguro}  These organic superconducting crystals are extremely brittle and have low critical temperatures.  Thus, there has been relatively little technological interest in organic superconductors to date.  The most prominent attempt to overcome the unattractive materials properties of organic charge transfer salts are the studies of microcrystals of $\beta$-(BEDT-TTF)$_2$I$_3$ embedded in a polycarbonate matrix.\cite{Tracz2001,Tracz1996,Jeszka1999,Laukhina1995} These composite materials retain many of the desirable materials properties of polycarbonate, such as its flexibility, and display some hints of superconductivity including a partial Meissner effect\cite{Tracz2001} and a drop in resistivity\cite{Tracz1996,Laukhina1995} below $\sim$5~K. However, there are no reports of such materials displaying zero electrical resistance.

It has been found that exposing a strongly insulating polymer to focused ion beams or metal plasmas can  increase its room temperature electrical conductivity by over 10 orders of magnitude, in some cases to as high as $\sim$10$^{3}$~S/cm, due to carbonization of the polymer by the ion beam.\cite{Forrest1982,Osaheni1992,Han07,HanJAPS,Powles} However, these materials remain insulators, exhibiting increasing (activated) resistivity with decreasing temperature. Achieving metallic conductivity in ion-implanted polymers is a long-standing problem. Recently, implantation of polyetheretherketone (PEEK) using a metallic Sn ion beam was explored.\cite{Tavenner2004} However, this resulted in a maximally implanted ion content insufficient for metallic conductivity due to  self-limiting sputtering processes. One way to overcome this problem is to deposit a thin metal layer on the polymer substrate and then use an ion beam to `mix' this metal into the polymer subsurface.\cite{Wang1997} Such `metal mixing' allows inert lower mass ions to be used, greatly reducing the sputtering. Nevertheless, the metal layer ensures that after implantation, large numbers of metal atoms have been mixed into the polymer. It has been shown that this process can give rise to metallic conductivity and even superconductivity.\cite{Micolich2006}

The reports, thus far, of superconductivity in metal mixed polymers are restricted to rather low temperatures (2-3~K). However the materials properties remain intriguing; most prominently, from a technological perspective, these metal mixed polymer superconductors retain the mechanical flexibility of the parent polymer.\cite{Micolich2006} Further, significant scientific questions still remain concerning the metallic and superconducting states in these systems. For example, the origin of the superconductivity has not yet been identified: is there a thin layer of metal below the surface of the polymer, a percolated network of metallic granules, or is the polymer-metal hybrid an intrinsically superconducting material? 

The electrical properties of these materials are certainly intriguing. While it has been shown that metal mixed superconducting polymers with a metallic normal state can be produced, their residual resistivity ratios (RRR) defined as $\rho(300~$K$)/\rho(T^{+}_{c})$, where $\rho(T^{+}_{c})$ is the resistance at a temperature slightly above the superconducting transition, are extremely small ($\sim1.2$), indicating that these are extremely disordered systems.\cite{Micolich2006} This is not unexpected given the manner in which these materials are produced. What is unexpected is the observation that both the critical temperature, $T_{c}$, and the critical field, $B_{c}$, are lower that than of the unimplanted metal film (compare Ref. \onlinecite{Micolich2006} with the results presented below). This is surprising for a thin film on metal on the surface of the plastic and lends weight to the possibility of more exotic explanations for the origin of the superconductivity.

If these materials prove to be tunable, as one naturally suspects they might, then they could serve as simple, cheap, experimental test beds for some of the most  profound questions about superconductivity in reduced dimensions including  superconductor-insulator transitions,\cite{Goldman2003,Markovic1999}  superconducting Kosterlitz-Thouless (KT) phase transitions,\cite{Kosterlitz1973} and superinsulation.\cite{Vinokur2008} Further, metal mixed polymer superconductors may prove to be excellent system in which to study percolated\cite{Yamada2004,Hua2007} and granular superconductivity\cite{Santos2006} and the competition between weak  localization and superconductivity.\cite{Oszwaldowski2002,Yamada2004} Finally, control of the substrate and/or implantation process could even allow for the controlled study of disorder in these systems.\cite{Myojin2007,disorder}

In order to begin to address the above scientific questions, and to move forward on possible technological and scientific applications, it is vital to have good control of the materials properties of the system. This control is required in two, quite separate, facets of preparation: (i) controlling the properties of the metal-polymer system prior to ion-implantation; and (ii) the ion-implantation process itself. Below we address (i) by reporting the results of a study of the electrical and optical properties of unimplanted thin films of an SnSb alloy on PEEK. These results will also provide a benchmark against which to examine properties of metal mixed polymers.

\section{Method}
Tin-antimony (SnSb) metallic thin films on polyetheretherketone (PEEK) were prepared and contacted in two different ways. Set A were made by evaporating a 95\%:5\% SnSb alloy ($\rho=7.28$~g~cm$^{-3}$) onto a 0.1~mm thick PEEK substrate (obtained from the Goodfellow Corporation). The substrate was cleaned with ethanol prior to deposition. The nominal thickness of the film was determined from a quartz crystal monitor located adjacent to the substrate during the vacuum deposition process. The metal was deposited at a maximum rate of 0.4~nm~s$^{-1}$. As an independent means of characterizing film thickness, absorbance spectra were taken of all the thin films using a dual beam Varian Cary 5000 UV-Vis-NIR spectrometer. A 1~mm diameter circular aperture was used to illuminate the samples. The samples were then rewashed in ethanol, and 2~mm wide gold contacts were deposited using a shadow mask and a similar vacuum evaporation process to that used for depositing SnSb. All evaporations were performed with a maximum initial pressure of $10^{-5}$~mbar. Wires were attached to the gold contacts using conducting silver paint (RS Components) with a conductivity of $\sigma=1000$~S/cm.

The DC electrical properties of set A were assessed using a 4-terminal measurement in a Hall bar configuration.\cite{VanderPauw1958b} The sample was mounted in a liquid nitrogen system (Oxford Instrument Optistat), and current-voltage (IV) sweeps were made over a range of temperatures between 77~K and 300~K. The sheet resistance can be determined from the measured resistance and the known measurement geometry. The current was sourced using a Agilent E3640A DC power supply and measured using a Keithley 6485 Picoammeter. The voltage was measured using a Keithley 2400 Source-Measure unit. 

Sample set B, used for determining the superconducting properties, had a different arrangement to the Hall bar described above. The samples in this second set were 15~mm square with 5~mm radius circular contacts deposited in the corners giving a quasi van der Pauw configuration.\cite{VanderPauw1958a} Copper wires were attached using InAg solder. Low temperature measurements were performed in an Oxford Instruments VTI system, which had a temperature range of 1.2 K to 200 K. The 2-terminal DC electrical resistance of the samples was measured using a Keithley 2400 Source-Measure unit.

\section{Results and Discussion}

 
\begin{figure*}


%
  \centering
		\includegraphics[width=0.48\textwidth]{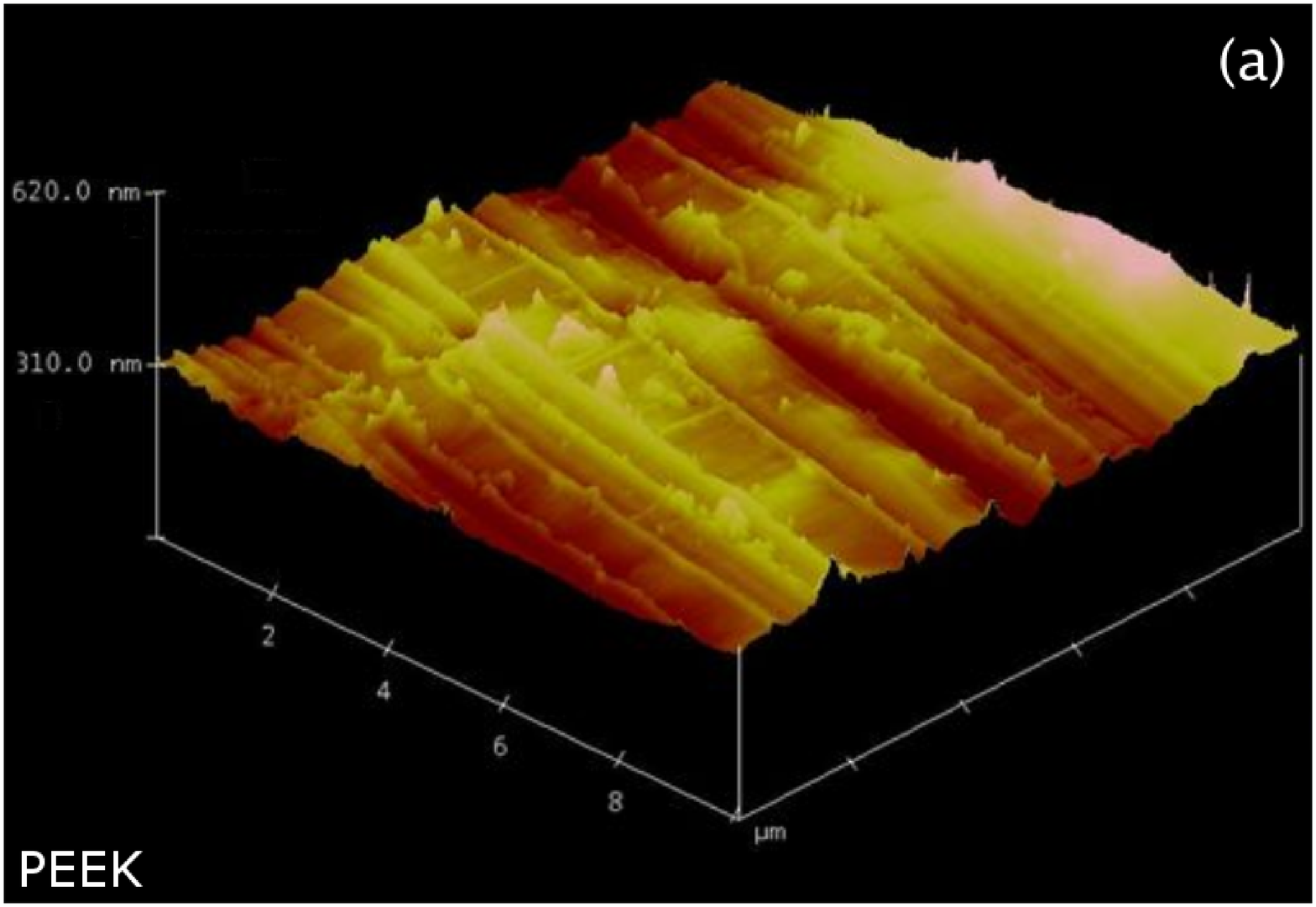}
		\includegraphics[width=0.48\textwidth]{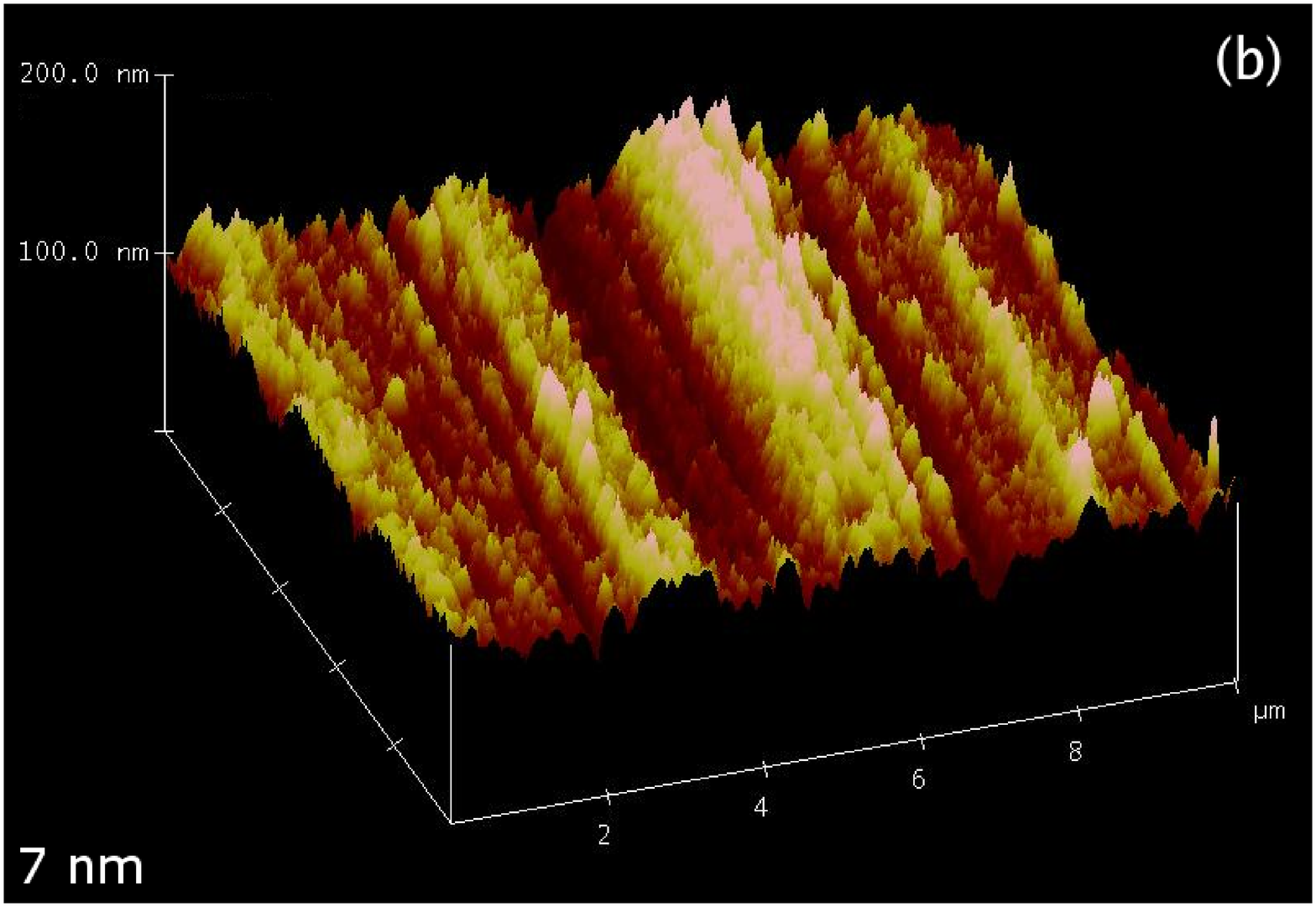}\\
		\includegraphics[width=0.48\textwidth]{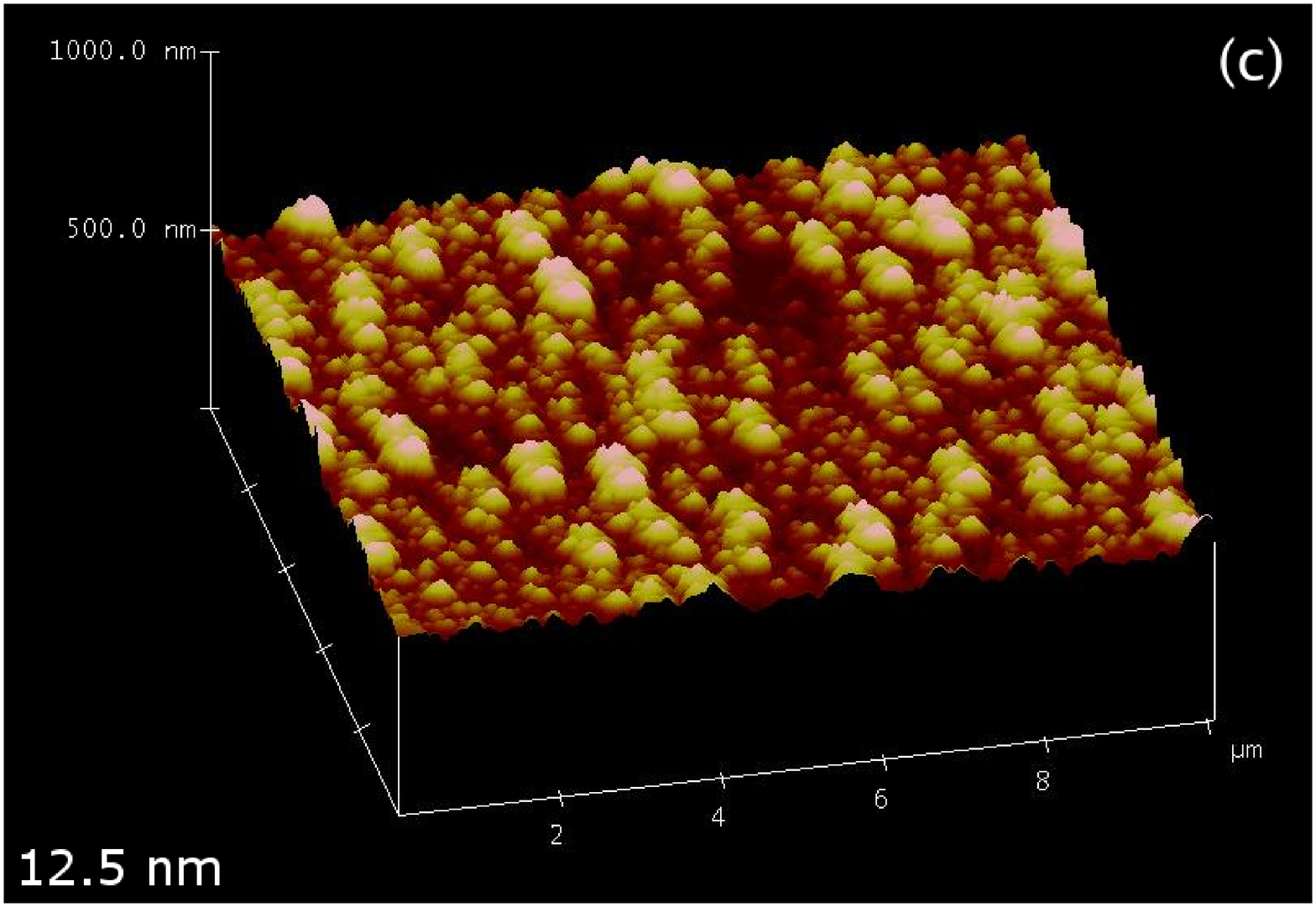}
		\includegraphics[width=0.48\textwidth]{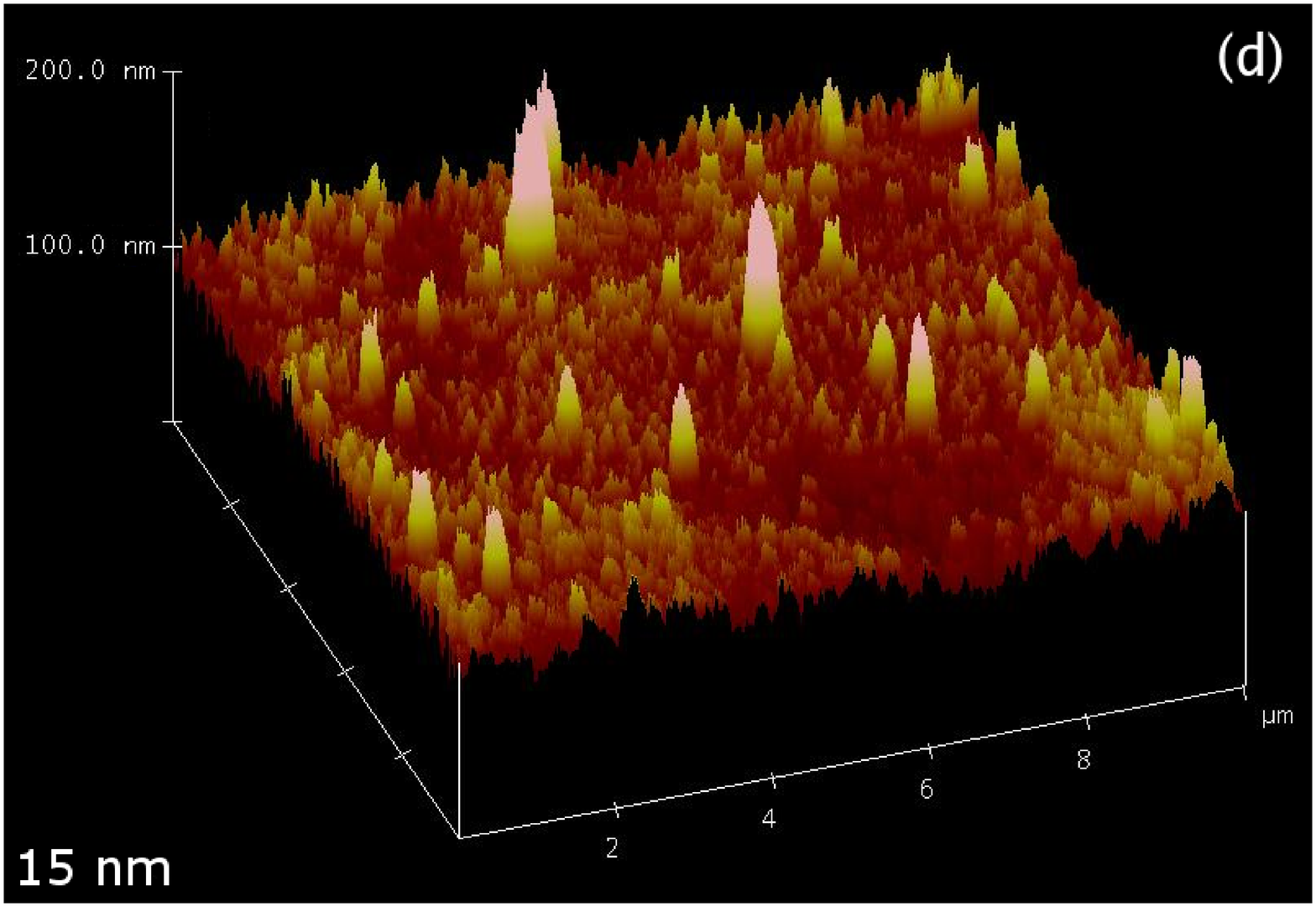}
\caption{Atomic force microscopy (AFM) images of (a) virgin PEEK surface and PEEK coated with SnSb thin films of thickness (b) 7.5~nm, (c) 12~nm and (d) 15~nm. The virgin PEEK surface is dominated by periodic striations $\sim1~\mu$m apart running parallel across the surface with a maximum height of 80~nm. As the film thickness is increased these striations are gradually filled, and have almost disappeared entirely once the thickness reaches $\sim$15~nm.} 

	\label{fig:002(3d)}
\end{figure*}

Figure \ref{fig:002(3d)}(a) shows an atomic force microscopy (AFM) image of the uncoated (native) polymer surface. It is very rough with prominent striations $\sim1~\mu$m apart and $\sim$80~nm high resulting from the PEEK manufacturing process (extrusion). It has recently been shown that such in-plane line defects can behave significantly differently from point defects in a thin film superconductor.\cite{Myojin2007} These striations dominate the morphology of very thin films , as is evident in Fig. \ref{fig:002(3d)}(b), which shows an AFM image of a 7~nm film. The presence of these striations raises the question of what impact they have on the superconductivity of thin films deposited upon them. Do these ridges act like line defects in a 2D film, or will they cause an asymmetry of the current flow for films whose morphology is dominated by that of the substrate? However, one should note that for films greater than 10~nm [Fig. \ref{fig:002(3d)}(c) and (d)] the striations no longer dominate the morphology and instead we see granular structures characteristic of the metallic film itself.


\begin{figure}
	\centering
		\includegraphics[width=0.48\textwidth]{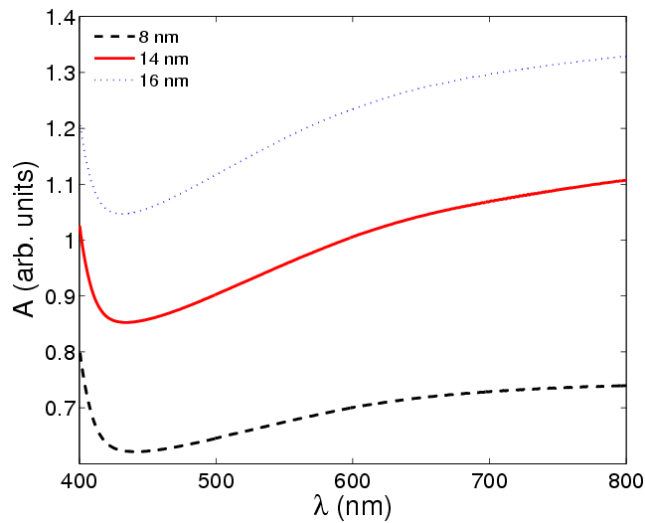}
	\caption{The absorbance spectra, $A(\lambda)$, for SnSb films of varying nominal thickness on PEEK substrates between $\lambda$ = 400~nm and 800~nm. For simplicity, we chose to characterize the film thickness by the 500nm absorbance value (note this choice is somewhat arbitrary since the spectral shape is smooth above 430 nm). We have repeated the analysis reported in Figures \ref{fig:AbsvThick500}~-~\ref{fig:convthickpar} with a number of other wavelengths (in the range 500-800 nm) and the results show no significant  differences.}
	\label{fig:AbsorbanceSpectrum}
\end{figure}


The crystal monitor is calibrated to give the correct thickness of metal evaporated onto a quartz substrate. Given the rather different wetting characteristics of PEEK and quartz, one does not expect the recorded absolute thickness to be an accurate measurement of the thickness of SnSb deposited on PEEK. As an independent means of characterizing the amount of metal deposited on the film, absorbance spectra were obtained.

\begin{figure}
	\centering
		\includegraphics[width=0.48\textwidth]{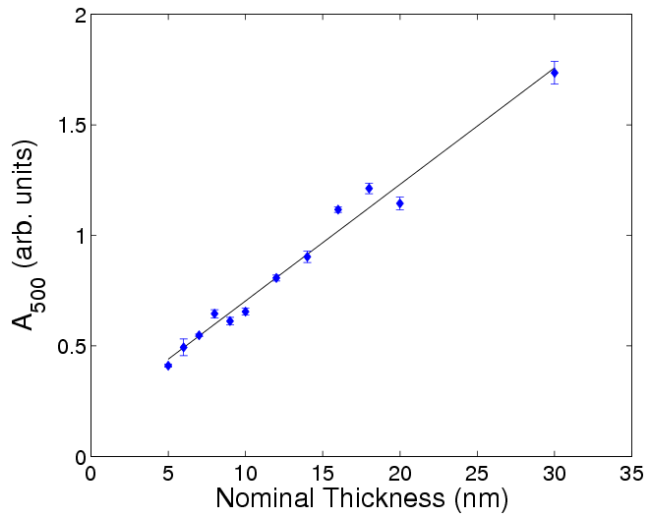}
	\caption{The relationship between absorbance at 500 nm, $A_{500}$, and the nominal thickness measured with a quartz crystal monitor of the metallic films. It is clear that there is a strong linear correlation between the absorbance and the thickness.}
	\label{fig:AbsvThick500}
\end{figure}

Figure~\ref{fig:AbsorbanceSpectrum} shows optical absorbance spectra, in the range $\lambda$=400-800~nm for 8, 14 and 16~nm SnSb films on PEEK. While only three thicknesses are shown in this figure, we have investigated a greater range and find qualitatively similar results, indicating that the absorbance might be a good alternative measurement of film thickness. To explore this, in Fig.~\ref{fig:AbsvThick500} we plot the absorbance at $\lambda$=~500~nm versus the nominal thickness, as measured by the quartz crystal monitor. There is a clear linear relationship between the optical absorbance and the nominal thickness recorded by the crystal monitor. This data suggests that the optical absorption at a fixed wavelength is at least as good a measure of the relative thickness of metal on unimplanted films as the crystal monitor. We will argue below that the absorbance is, in fact, a more reliable measurement of the relative thickness than the crystal monitor.

\begin{figure}
	\centering
		\includegraphics[width=0.48\textwidth]{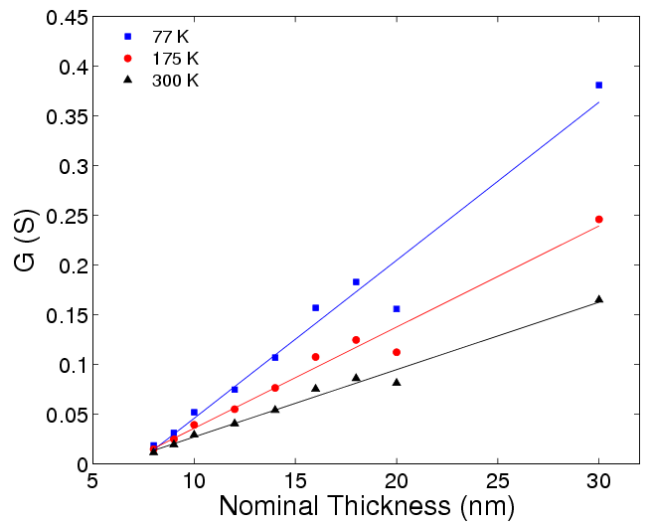}
	\caption{Sheet conductance, $G$, versus the nominal thickness of a tin/antimony (SnSb) metal film on a plastic (PEEK) substrate. The nominal thickness was taken as the value recorded by a quartz crystal monitor positioned next to the plastic substrate during metal deposition. Conductance data was obtained with the current flowing parallel to the striations of the substrate. 
The conductance of the samples increases with the amount of metal deposited. Note the anomalously small conductivity of the 20~nm sample.}
	\label{fig:ConvThickPar}
\end{figure}

\begin{figure}
	\centering
		\includegraphics[width=0.48\textwidth]{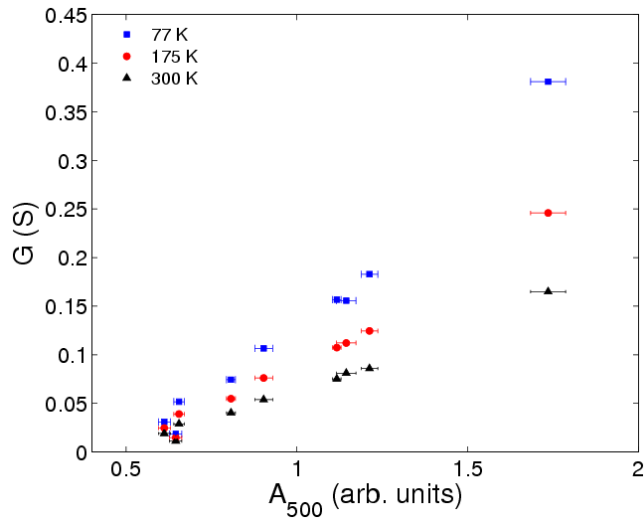}
	\caption{Sheet conductance, $G$, versus optical absorbance at 500nm, $A_{500}$, for a SnSb film on a PEEK substrate at temperatures ranging from 77~K to 300~K. Conductance data was taken with the current flowing parallel to the striations of the substrate. The anomaly seen in Fig.~\ref{fig:ConvThickPar} for the sample with a nominal thickness of 20~nm is absent. This suggests that the absorbance is a more reliable calibration of the amount of metal evaporated onto the PEEK substrate compared to the quartz crystal monitor. Further, the absolute values are not meaningful, as they correspond to the thickness of SnSb on quartz rather than PEEK. The same conclusion can be reached by studying data for variation of the conductance perpendicular to the striations with nominal thickness (not shown) and absorbance (Fig. \ref{fig:ConvAbsPer500}).}
	\label{fig:ConvAbsPar500}
\end{figure}

Figure~\ref{fig:ConvThickPar} shows the relationship between the sheet conductance and the nominal thickness of the SnSb film (i.e., the thickness measured by the quartz crystal monitor) at temperatures between 77 and 300~K. The conductance increases with nominal thickness as one would expect. The data is very smooth with the exception of an anomaly at 20~nm. For comparison, the relationship between the sheet conductance and optical absorbance is shown in Fig.~\ref{fig:ConvAbsPar500}. This data is also smooth but the anomaly at 20~nm in Fig.~\ref{fig:ConvThickPar} is now absent. This suggests that the absorbance provides a better characterization of the actual thickness of the metal on the plastic substrate than the nominal thickness recorded by the crystal monitor.

\begin{figure}
	\centering
		\includegraphics[width=0.48\textwidth]{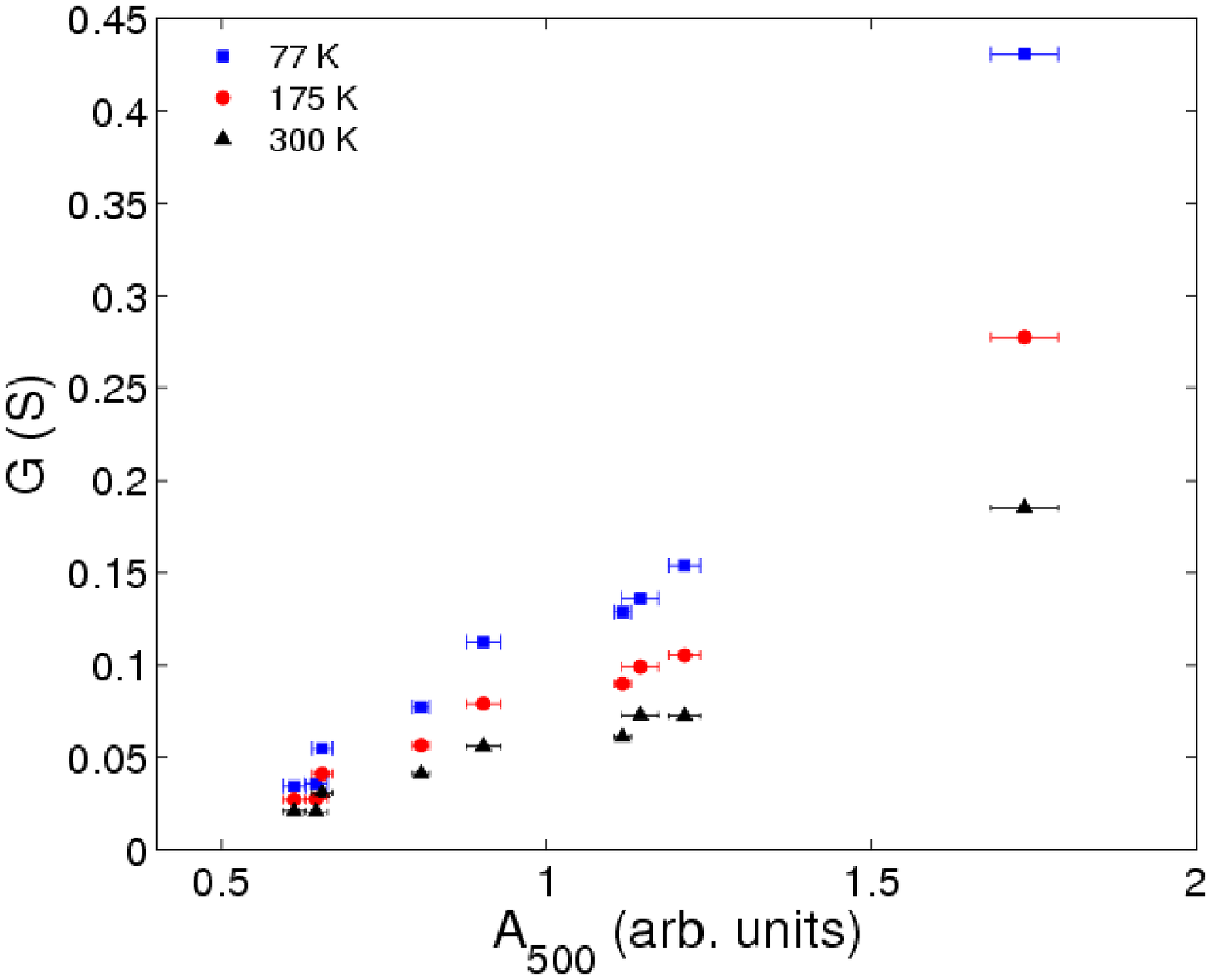}
	\caption{Similar data to that reported in Fig. \ref{fig:ConvAbsPar500}, but with the current flowing perpendicular to the striations. No appreciable differences are observed at any thickness (absorbance). Therefore we conclude that the striations do not have a significant effect on conductance of the films in the metal state. This is particularly surprising for the thinnest films where the striations dictate the morphology of the metal.}
	\label{fig:ConvAbsPer500}
\end{figure}

We have measured the conductivity both parallel and perpendicular to the striations to study their effect. This data is shown in Figs. \ref{fig:ConvAbsPar500} and \ref{fig:ConvAbsPer500} respectively. The morphology of the thinnest films is dominated by the striations of the substrate, as shown in Fig.~\ref{fig:002(3d)}. We were therefore surprised to discover no appreciable differences between the conductance measured at any thickness in either direction. To further test this result we calculated the gradient of the conductance versus absorbance relation for thicknesses less than 20~nm. This was repeated for $\lambda=$~500~nm, 600~nm, 700~nm and 800~nm. There was no significant difference in the calculated values. We repeated this analysis considering only the thinnest films (nominal thicknesses $\leq12$~nm); again, no difference was found, indicating that the striations have no measurable effect on the electrical properties of the SnSb films. This suggests that, in spite of the striations, the film is reasonably uniformly deposited on the surface of the PEEK, i.e., the metallic film is continuous and conformal. 

Extrapolating the data in Figs. \ref{fig:ConvAbsPar500} and \ref{fig:ConvAbsPer500} indicates that the conductance goes to zero at an absorbance corresponding to a nominal thickness of approximately 7~nm for both current orientations. Measurements were made on films with nominal thicknesses of 5~nm, 6~nm and 7~nm. These samples were found to be insulating with a resistance several orders of magnitude higher than those of the 8~nm samples. Again, this is consistent with the conclusion that the morphology of the substrate does not affect the electrical properties of the metal thin film. 

However, although we have shown that the striations do not strongly effect the deposition of metal on PEEK, their effect on the implantation process remains to be studied. We will report on this in a future publication. 

\begin{figure}
	\centering
		\includegraphics[width=0.48\textwidth]{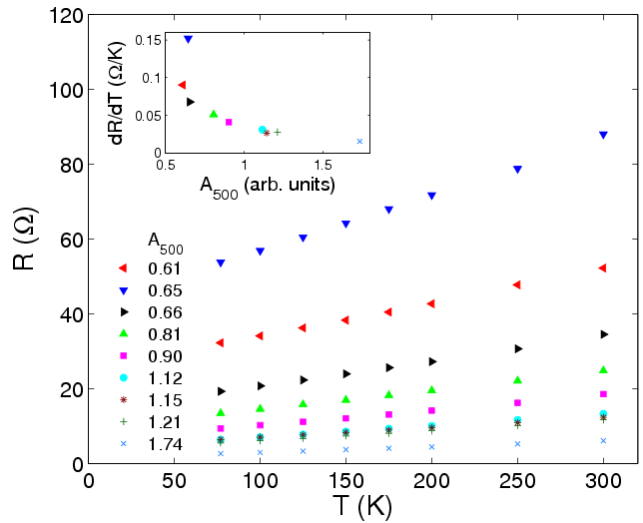}
		\includegraphics[width=0.48\textwidth]{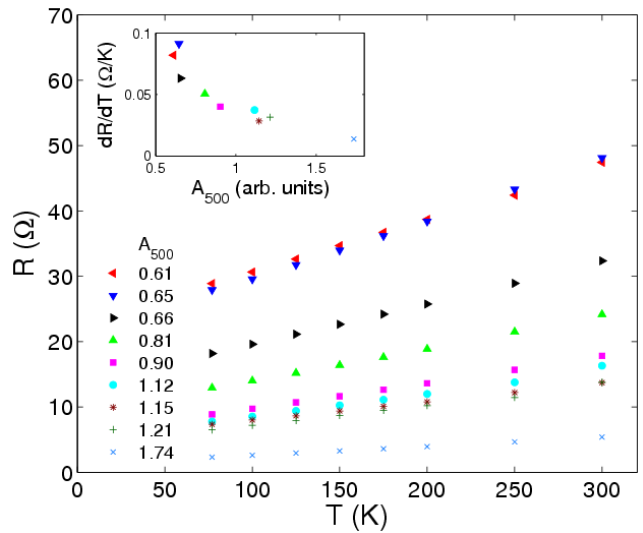}
	\caption{Resistance, $R$, versus temperature, $T$, profiles for SnSb films on PEEK substrates. The top panel shows data for current flowing parallel to the striations and the bottom panel shows data for current flowing perpendicular to the striations. It is evident that the resistance of the samples increases with temperature, indicating that the thin films are metallic.}
	\label{fig:convthickpar}
\end{figure}

The temperature dependence of the resistance is shown in Fig.~\ref{fig:convthickpar} for samples with metal layers $\geq8$~nm for temperatures ranging between 77~K and 300~K. The resistance monotonically increases with temperature, indicating that these samples are metallic. The consistency of the data indicates that, despite the relatively basic production process, the quality of these samples is quite high. It is interesting to note that gradient of the resistivity increases as the nominal thickness approaches 7~nm. It is rather puzzling that the gradient increases as the sample's resistance increases, i.e., as the metal layer becomes thinner.

\begin{figure}
	\centering
		\includegraphics[width=0.48\textwidth]{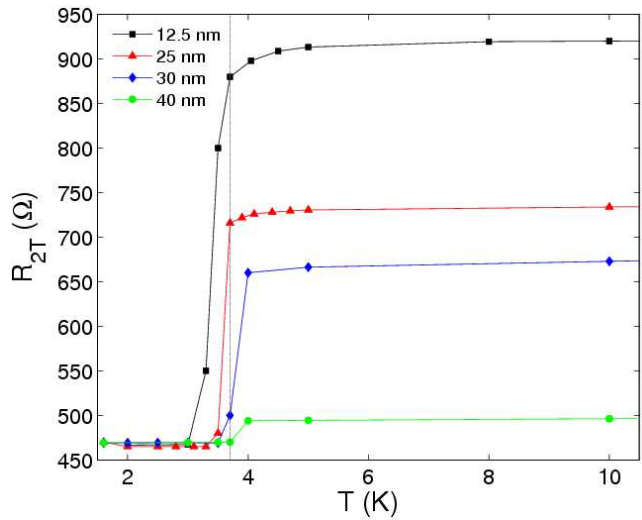}
	\caption{Temperature dependence of the two-terminal resistance of SnSb films on PEEK substrates between 1.5 - 10.5~K. The sharp drop in resistance indicates a superconducting phase transition. The onset critical temperature, $T_c$, does not seem to depend on the film thickness, but the transition is significantly broadened. This is in marked contrast to the transition in the metal mixed materials,\cite{Micolich2006} where $T_c$ is significantly suppressed. The $T_c=3.7$~K for bulk Sn is indicated by the vertical line.}
	\label{fig:R2TvTemp}
\end{figure}

We now turn to a study of the superconducting properties of our samples. This is important for benchmarking the superconducting properties of the metal mixed samples.\cite{Micolich2006} Current-voltage sweeps were obtained at temperatures down to 1.5~K. Figure \ref{fig:R2TvTemp} shows shows the temperature dependence of the two terminal resistance between 1.5 and 10.5~K for samples ranging in thickness between 12.5~nm and 40~nm, the former having the highest residual resistivity. The superconducting transition is significantly broadened, but the onset $T_c$ is \emph{not} suppressed by the decreased film thickness. This is in marked contrast to metal mixed polymer superconductors,\cite{Micolich2006} where a strong suppression of $T_c$ from that of bulk Sn (3.7~K) is observed.\cite{footSbsuper} This suggests that the physics of the superconducting state is significantly changed by the metal mixing process.

\section{Conclusion}

Metal mixed polymers are a cheap and effective way to produce flexible metals and superconductors.\cite{Micolich2006} As part of an on-going effort to learn how to tune the properties of these systems we have performed studies on the electrical properties of these systems prior to metal mixing. We have shown that the electrical properties of metallic thin films are remarkably robust to variations in the substrate morphology. We have also demonstrated that the optical absorbance of the films at a fixed wavelength provides a reliable and reproducible characterization of the relative film thickness. Additionally we have found that as the film thickness is reduced, the superconducting transition in the unimplanted thin films is broadened, but the onset of the transition remains at $\sim$3.7~K, the transition temperature of bulk Sn. This shows that the metal mixing process has a significant effect of the physics of the superconducting state beyond that achieved by simply reducing the film thickness.

\section{Acknowledgments}

We acknowledge insightful discussions with Alex Hamilton and Eddy Yusuf.
This work was supported by the Australian Research Council (ARC) under the Discovery scheme (project number DP0559215). PM acknowledges the Queensland Government for the funding through the Smart State Senior Fellowship Program. BJP is the recipient of an ARC Queen Elizabeth II Fellowship (project number DP0878523).













\begin{thebibliography}{31}
\expandafter\ifx\csname natexlab\endcsname\relax\def\natexlab#1{#1}\fi
\expandafter\ifx\csname bibnamefont\endcsname\relax
  \def\bibnamefont#1{#1}\fi
\expandafter\ifx\csname bibfnamefont\endcsname\relax
  \def\bibfnamefont#1{#1}\fi
\expandafter\ifx\csname citenamefont\endcsname\relax
  \def\citenamefont#1{#1}\fi
\expandafter\ifx\csname url\endcsname\relax
  \def\url#1{\texttt{#1}}\fi
\expandafter\ifx\csname urlprefix\endcsname\relax\def\urlprefix{URL }\fi
\providecommand{\bibinfo}[2]{#2}
\providecommand{\eprint}[2][]{\url{#2}}

\bibitem[{\citenamefont{{A. J. Heeger}}(2001)}]{HeegerRMP}
\bibinfo{author}{\bibnamefont{{A. J. Heeger}}}, \bibinfo{journal}{Rev. Mod.
  Phys.} \textbf{\bibinfo{volume}{73}}, \bibinfo{pages}{681}
  (\bibinfo{year}{2001}).

\bibitem[{\citenamefont{{A. G. MacDiarmid}}(2001)}]{MacDiarmidRMP}
\bibinfo{author}{\bibnamefont{{A. G. MacDiarmid}}}, \bibinfo{journal}{Rev. Mod.
  Phys.} \textbf{\bibinfo{volume}{73}}, \bibinfo{pages}{701}
  (\bibinfo{year}{2001}).

\bibitem[{\citenamefont{{H. Shirakawa}}(2001)}]{ShirakawaRMP}
\bibinfo{author}{\bibnamefont{{H. Shirakawa}}}, \bibinfo{journal}{Rev. Mod.
  Phys.} \textbf{\bibinfo{volume}{73}}, \bibinfo{pages}{713}
  (\bibinfo{year}{2001}).

\bibitem[{\citenamefont{Voss}(2000)}]{Voss2000}
\bibinfo{author}{\bibfnamefont{D.}~\bibnamefont{Voss}},
  \bibinfo{journal}{Nature} \textbf{\bibinfo{volume}{407}},
  \bibinfo{pages}{442} (\bibinfo{year}{2000}).

\bibitem[{\citenamefont{{For a recent review, see, B. J. Powell and R. H.
  McKenzie}}(2006)}]{Powell2006}
\bibinfo{author}{\bibnamefont{{For a recent review, see, B. J. Powell and R. H.
  McKenzie}}}, \bibinfo{journal}{J. Phys. Condens. Matter}
  \textbf{\bibinfo{volume}{18}}, \bibinfo{pages}{R827} (\bibinfo{year}{2006}).

\bibitem[{\citenamefont{Ishiguro et~al.}(2001)\citenamefont{Ishiguro, Yamaji,
  and Saito}}]{Ishiguro}
\bibinfo{author}{\bibfnamefont{T.}~\bibnamefont{Ishiguro}},
  \bibinfo{author}{\bibfnamefont{K.}~\bibnamefont{Yamaji}}, \bibnamefont{and}
  \bibinfo{author}{\bibfnamefont{G.}~\bibnamefont{Saito}},
  \emph{\bibinfo{title}{Organic Superconductors}}
  (\bibinfo{publisher}{Springer, Berlin}, \bibinfo{year}{2001}).

\bibitem[{\citenamefont{Tracz et~al.}(2001)\citenamefont{Tracz, Wosnitza,
  Barakat, Hagel, and Muller}}]{Tracz2001}
\bibinfo{author}{\bibfnamefont{A.}~\bibnamefont{Tracz}},
  \bibinfo{author}{\bibfnamefont{J.}~\bibnamefont{Wosnitza}},
  \bibinfo{author}{\bibfnamefont{S.}~\bibnamefont{Barakat}},
  \bibinfo{author}{\bibfnamefont{J.}~\bibnamefont{Hagel}}, \bibnamefont{and}
  \bibinfo{author}{\bibfnamefont{H.}~\bibnamefont{Muller}},
  \bibinfo{journal}{Synth. Met.} \textbf{\bibinfo{volume}{120}},
  \bibinfo{pages}{849} (\bibinfo{year}{2001}).

\bibitem[{\citenamefont{Tracz et~al.}(1996)\citenamefont{Tracz, Jeszka, and
  Sroczyriska}}]{Tracz1996}
\bibinfo{author}{\bibfnamefont{A.}~\bibnamefont{Tracz}},
  \bibinfo{author}{\bibfnamefont{J.~K.} \bibnamefont{Jeszka}},
  \bibnamefont{and}
  \bibinfo{author}{\bibfnamefont{A.}~\bibnamefont{Sroczyriska}},
  \bibinfo{journal}{Adv. Mat. Opt. Elect.} \textbf{\bibinfo{volume}{6}},
  \bibinfo{pages}{335} (\bibinfo{year}{1996}).

\bibitem[{\citenamefont{Jeszka et~al.}(1999)\citenamefont{Jeszka, Tracz,
  Scrozynska, Ulanski, Muller, Pakula, and Kryszewski}}]{Jeszka1999}
\bibinfo{author}{\bibfnamefont{J.~K.} \bibnamefont{Jeszka}},
  \bibinfo{author}{\bibfnamefont{A.}~\bibnamefont{Tracz}},
  \bibinfo{author}{\bibfnamefont{A.}~\bibnamefont{Scrozynska}},
  \bibinfo{author}{\bibfnamefont{J.}~\bibnamefont{Ulanski}},
  \bibinfo{author}{\bibfnamefont{H.}~\bibnamefont{Muller}},
  \bibinfo{author}{\bibfnamefont{T.}~\bibnamefont{Pakula}}, \bibnamefont{and}
  \bibinfo{author}{\bibfnamefont{M.}~\bibnamefont{Kryszewski}},
  \bibinfo{journal}{Synth. Met.} \textbf{\bibinfo{volume}{103}},
  \bibinfo{pages}{1820} (\bibinfo{year}{1999}).

\bibitem[{\citenamefont{Laukhina et~al.}(1995)\citenamefont{Laukhina,
  Merzhanov, Pesotskii, Khomenko, Yagubskii, Ulanski, and
  Jeszka}}]{Laukhina1995}
\bibinfo{author}{\bibfnamefont{E.~E.} \bibnamefont{Laukhina}},
  \bibinfo{author}{\bibfnamefont{V.~A.} \bibnamefont{Merzhanov}},
  \bibinfo{author}{\bibfnamefont{S.~I.} \bibnamefont{Pesotskii}},
  \bibinfo{author}{\bibfnamefont{A.~G.} \bibnamefont{Khomenko}},
  \bibinfo{author}{\bibfnamefont{E.~B.} \bibnamefont{Yagubskii}},
  \bibinfo{author}{\bibfnamefont{J.}~\bibnamefont{Ulanski}}, \bibnamefont{and}
  \bibinfo{author}{\bibfnamefont{J.~K.} \bibnamefont{Jeszka}},
  \bibinfo{journal}{Synth. Met.} \textbf{\bibinfo{volume}{70}},
  \bibinfo{pages}{797} (\bibinfo{year}{1995}).

\bibitem[{\citenamefont{Forrest et~al.}(1982)\citenamefont{Forrest, Kaplan,
  Sshmidt, Venkatesan, and Lovinger}}]{Forrest1982}
\bibinfo{author}{\bibfnamefont{S.~R.} \bibnamefont{Forrest}},
  \bibinfo{author}{\bibfnamefont{M.~L.} \bibnamefont{Kaplan}},
  \bibinfo{author}{\bibfnamefont{P.~H.} \bibnamefont{Sshmidt}},
  \bibinfo{author}{\bibfnamefont{T.}~\bibnamefont{Venkatesan}},
  \bibnamefont{and} \bibinfo{author}{\bibfnamefont{A.~J.}
  \bibnamefont{Lovinger}}, \bibinfo{journal}{Appl. Phys. Lett.}
  \textbf{\bibinfo{volume}{41}}, \bibinfo{pages}{708} (\bibinfo{year}{1982}).

\bibitem[{\citenamefont{Osaheni et~al.}(1992)\citenamefont{Osaheni, Jenekhe,
  Burns, Du, Joo, Epstein, and Wang}}]{Osaheni1992}
\bibinfo{author}{\bibfnamefont{J.~A.} \bibnamefont{Osaheni}},
  \bibinfo{author}{\bibfnamefont{S.~A.} \bibnamefont{Jenekhe}},
  \bibinfo{author}{\bibfnamefont{A.}~\bibnamefont{Burns}},
  \bibinfo{author}{\bibfnamefont{G.}~\bibnamefont{Du}},
  \bibinfo{author}{\bibfnamefont{J.}~\bibnamefont{Joo}},
  \bibinfo{author}{\bibfnamefont{A.~J.} \bibnamefont{Epstein}},
  \bibnamefont{and} \bibinfo{author}{\bibfnamefont{C.~S.} \bibnamefont{Wang}},
  \bibinfo{journal}{Macromolecules} \textbf{\bibinfo{volume}{25}},
  \bibinfo{pages}{5828} (\bibinfo{year}{1992}).

\bibitem[{\citenamefont{{Z. J. Han, B. K. Tay, P. C. T. Ha, M. Shakerzadeh, A.
  A. Cimmino, S. Prawer, and D. McKenzie}}(2007)}]{Han07}
\bibinfo{author}{\bibnamefont{{Z. J. Han, B. K. Tay, P. C. T. Ha, M.
  Shakerzadeh, A. A. Cimmino, S. Prawer, and D. McKenzie}}},
  \bibinfo{journal}{Appl. Phys. Lett.} \textbf{\bibinfo{volume}{91}},
  \bibinfo{pages}{052103} (\bibinfo{year}{2007}).

\bibitem[{\citenamefont{{Z. J. Han and B. K. Tay}}(2007)}]{HanJAPS}
\bibinfo{author}{\bibnamefont{{Z. J. Han and B. K. Tay}}}, \bibinfo{journal}{J.
  Appl. Polym. Sci.} \textbf{\bibinfo{volume}{107}}, \bibinfo{pages}{3332}
  (\bibinfo{year}{2007}).

\bibitem[{\citenamefont{{R.C. Powles, D.R. McKenzie, N. Fujisawa, D.G.
  McCulloch}}(2005)}]{Powles}
\bibinfo{author}{\bibnamefont{{R.C. Powles, D.R. McKenzie, N. Fujisawa, D.G.
  McCulloch}}}, \bibinfo{journal}{Diam. Rel. Mat.}
  \textbf{\bibinfo{volume}{14}}, \bibinfo{pages}{1577} (\bibinfo{year}{2005}).

\bibitem[{\citenamefont{Tavenner et~al.}(2004)\citenamefont{Tavenner, Meredith,
  Wood, Curry, and Giedd}}]{Tavenner2004}
\bibinfo{author}{\bibfnamefont{E.}~\bibnamefont{Tavenner}},
  \bibinfo{author}{\bibfnamefont{P.}~\bibnamefont{Meredith}},
  \bibinfo{author}{\bibfnamefont{B.}~\bibnamefont{Wood}},
  \bibinfo{author}{\bibfnamefont{M.}~\bibnamefont{Curry}}, \bibnamefont{and}
  \bibinfo{author}{\bibfnamefont{R.}~\bibnamefont{Giedd}},
  \bibinfo{journal}{Synth. Met.} \textbf{\bibinfo{volume}{145}},
  \bibinfo{pages}{183} (\bibinfo{year}{2004}).

\bibitem[{\citenamefont{Wang et~al.}(1997)\citenamefont{Wang, Giedd, Moss, and
  Kaufmann}}]{Wang1997}
\bibinfo{author}{\bibfnamefont{Y.~Q.} \bibnamefont{Wang}},
  \bibinfo{author}{\bibfnamefont{R.~E.} \bibnamefont{Giedd}},
  \bibinfo{author}{\bibfnamefont{M.~G.} \bibnamefont{Moss}}, \bibnamefont{and}
  \bibinfo{author}{\bibfnamefont{J.}~\bibnamefont{Kaufmann}},
  \bibinfo{journal}{Nucl. Instr. Meth. B} \textbf{\bibinfo{volume}{127/128}},
  \bibinfo{pages}{710} (\bibinfo{year}{1997}).

\bibitem[{\citenamefont{Micolich et~al.}(2006)\citenamefont{Micolich, Tavenner,
  Powell, Hamilton, Curry, Giedd, and Meredith}}]{Micolich2006}
\bibinfo{author}{\bibfnamefont{A.~P.} \bibnamefont{Micolich}},
  \bibinfo{author}{\bibfnamefont{E.}~\bibnamefont{Tavenner}},
  \bibinfo{author}{\bibfnamefont{B.~J.} \bibnamefont{Powell}},
  \bibinfo{author}{\bibfnamefont{A.~R.} \bibnamefont{Hamilton}},
  \bibinfo{author}{\bibfnamefont{M.~T.} \bibnamefont{Curry}},
  \bibinfo{author}{\bibfnamefont{R.~E.} \bibnamefont{Giedd}}, \bibnamefont{and}
  \bibinfo{author}{\bibfnamefont{P.}~\bibnamefont{Meredith}},
  \bibinfo{journal}{Appl. Phys. Lett.} \textbf{\bibinfo{volume}{89}},
  \bibinfo{pages}{152503} (\bibinfo{year}{2006}).

\bibitem[{\citenamefont{Goldman}(2003)}]{Goldman2003}
\bibinfo{author}{\bibfnamefont{A.~M.} \bibnamefont{Goldman}},
  \bibinfo{journal}{Physica E} \textbf{\bibinfo{volume}{18}},
  \bibinfo{pages}{1} (\bibinfo{year}{2003}).

\bibitem[{\citenamefont{Markovic et~al.}(1999)\citenamefont{Markovic,
  Christiansen, Mack, Huber, and Goldman}}]{Markovic1999}
\bibinfo{author}{\bibfnamefont{N.}~\bibnamefont{Markovic}},
  \bibinfo{author}{\bibfnamefont{C.}~\bibnamefont{Christiansen}},
  \bibinfo{author}{\bibfnamefont{A.~M.} \bibnamefont{Mack}},
  \bibinfo{author}{\bibfnamefont{W.~H.} \bibnamefont{Huber}}, \bibnamefont{and}
  \bibinfo{author}{\bibfnamefont{A.~M.} \bibnamefont{Goldman}},
  \bibinfo{journal}{Phys. Rev. B} \textbf{\bibinfo{volume}{60}},
  \bibinfo{pages}{4320} (\bibinfo{year}{1999}).

\bibitem[{\citenamefont{Kosterlitz and Thouless}(1973)}]{Kosterlitz1973}
\bibinfo{author}{\bibfnamefont{J.~M.} \bibnamefont{Kosterlitz}}
  \bibnamefont{and} \bibinfo{author}{\bibfnamefont{D.~J.}
  \bibnamefont{Thouless}}, \bibinfo{journal}{J. Phys. C}
  \textbf{\bibinfo{volume}{6}}, \bibinfo{pages}{1181} (\bibinfo{year}{1973}).

\bibitem[{\citenamefont{Vinokur et~al.}(2008)\citenamefont{Vinokur, Baturina,
  Fistul, Mironov, Baklanov, and Strunk}}]{Vinokur2008}
\bibinfo{author}{\bibfnamefont{V.~M.} \bibnamefont{Vinokur}},
  \bibinfo{author}{\bibfnamefont{T.~I.} \bibnamefont{Baturina}},
  \bibinfo{author}{\bibfnamefont{M.~V.} \bibnamefont{Fistul}},
  \bibinfo{author}{\bibfnamefont{A.~Y.} \bibnamefont{Mironov}},
  \bibinfo{author}{\bibfnamefont{M.~R.} \bibnamefont{Baklanov}},
  \bibnamefont{and} \bibinfo{author}{\bibfnamefont{C.}~\bibnamefont{Strunk}},
  \bibinfo{journal}{Nature} \textbf{\bibinfo{volume}{452}},
  \bibinfo{pages}{613} (\bibinfo{year}{2008}).

\bibitem[{\citenamefont{Yamada et~al.}(2004)\citenamefont{Yamada, Shinozaki,
  and Kawaguti}}]{Yamada2004}
\bibinfo{author}{\bibfnamefont{K.}~\bibnamefont{Yamada}},
  \bibinfo{author}{\bibfnamefont{B.}~\bibnamefont{Shinozaki}},
  \bibnamefont{and} \bibinfo{author}{\bibfnamefont{T.}~\bibnamefont{Kawaguti}},
  \bibinfo{journal}{Phys. Rev. B} \textbf{\bibinfo{volume}{70}},
  \bibinfo{pages}{144503} (\bibinfo{year}{2004}).

\bibitem[{\citenamefont{Hua et~al.}(2007)\citenamefont{Hua, Xiao, Rosenmann,
  Beloborodov, Welp, Kwop, and Crabtree}}]{Hua2007}
\bibinfo{author}{\bibfnamefont{J.}~\bibnamefont{Hua}},
  \bibinfo{author}{\bibfnamefont{Z.~L.} \bibnamefont{Xiao}},
  \bibinfo{author}{\bibfnamefont{D.}~\bibnamefont{Rosenmann}},
  \bibinfo{author}{\bibfnamefont{C.}~\bibnamefont{Beloborodov}},
  \bibinfo{author}{\bibfnamefont{U.}~\bibnamefont{Welp}},
  \bibinfo{author}{\bibfnamefont{W.}~\bibnamefont{Kwop}}, \bibnamefont{and}
  \bibinfo{author}{\bibfnamefont{G.~W.} \bibnamefont{Crabtree}},
  \bibinfo{journal}{Appl. Phys. Lett.} \textbf{\bibinfo{volume}{90}},
  \bibinfo{pages}{072507} (\bibinfo{year}{2007}).

\bibitem[{\citenamefont{dos Santos et~al.}(2006)\citenamefont{dos Santos,
  Oliveira, da~Luz, Bortolozo, Sandim, and Machado}}]{Santos2006}
\bibinfo{author}{\bibfnamefont{C.~A.~M.} \bibnamefont{dos Santos}},
  \bibinfo{author}{\bibfnamefont{C.~J.~V.} \bibnamefont{Oliveira}},
  \bibinfo{author}{\bibfnamefont{M.~S.} \bibnamefont{da~Luz}},
  \bibinfo{author}{\bibfnamefont{A.~D.} \bibnamefont{Bortolozo}},
  \bibinfo{author}{\bibfnamefont{M.~J.~R.} \bibnamefont{Sandim}},
  \bibnamefont{and} \bibinfo{author}{\bibfnamefont{A.~J.~S.}
  \bibnamefont{Machado}}, \bibinfo{journal}{Phys. Rev. B}
  \textbf{\bibinfo{volume}{74}}, \bibinfo{pages}{184526}
  (\bibinfo{year}{2006}).

\bibitem[{\citenamefont{Oszwaldowski et~al.}(2002)\citenamefont{Oszwaldowski,
  Berus, and Dugaev}}]{Oszwaldowski2002}
\bibinfo{author}{\bibfnamefont{M.}~\bibnamefont{Oszwaldowski}},
  \bibinfo{author}{\bibfnamefont{T.}~\bibnamefont{Berus}}, \bibnamefont{and}
  \bibinfo{author}{\bibfnamefont{V.~K.} \bibnamefont{Dugaev}},
  \bibinfo{journal}{Phys. Rev. B} \textbf{\bibinfo{volume}{65}},
  \bibinfo{pages}{235418} (\bibinfo{year}{2002}).

\bibitem[{\citenamefont{Myojin and Ikeda}(2007)}]{Myojin2007}
\bibinfo{author}{\bibfnamefont{K.}~\bibnamefont{Myojin}} \bibnamefont{and}
  \bibinfo{author}{\bibfnamefont{R.}~\bibnamefont{Ikeda}}, \bibinfo{journal}{J.
  Phys. Soc. Jpn} \textbf{\bibinfo{volume}{76}}, \bibinfo{pages}{094710}
  (\bibinfo{year}{2007}).

\bibitem[{\citenamefont{{B. J. Powell and R. H. McKenzie}}(2004)}]{disorder}
\bibinfo{author}{\bibnamefont{{B. J. Powell and R. H. McKenzie}}},
  \bibinfo{journal}{Phys. Rev. B} \textbf{\bibinfo{volume}{69}},
  \bibinfo{pages}{024519} (\bibinfo{year}{2004}).

\bibitem[{\citenamefont{Van~der Pauw}(1958{\natexlab{a}})}]{VanderPauw1958b}
\bibinfo{author}{\bibfnamefont{L.}~\bibnamefont{Van~der Pauw}},
  \bibinfo{journal}{Philips Technical Review} \textbf{\bibinfo{volume}{26}},
  \bibinfo{pages}{220} (\bibinfo{year}{1958}{\natexlab{a}}).

\bibitem[{\citenamefont{Van~der Pauw}(1958{\natexlab{b}})}]{VanderPauw1958a}
\bibinfo{author}{\bibfnamefont{L.}~\bibnamefont{Van~der Pauw}},
  \bibinfo{journal}{Philips Res. Repts.} \textbf{\bibinfo{volume}{13}},
  \bibinfo{pages}{1} (\bibinfo{year}{1958}{\natexlab{b}}).

\bibitem[{foo()}]{footSbsuper}
\bibinfo{howpublished}{{Note that the small amount of Sb in the alloy
  stabilizes the metallic, white, phase of Sn, but, otherwise, does not
  significantly affect the superconducting properties.}}

\end{thebibliography}
\end{document}